\documentclass[prl,twocolumn,aps,showpacs]{revtex4}
\usepackage{epsfig}
\usepackage{bm}
\begin{document}
\title{Pairing and condensation in a resonant 
              Bose-Fermi mixture}
\author{Elisa Fratini and Pierbiagio Pieri}
\affiliation{Dipartimento di Fisica, Scuola di Scienze e Tecnologie,
Universit\`{a} di Camerino, Via Madonna delle Carceri 9, I-62032 Camerino, Italy}
\date{\today}

\begin{abstract}
We study by diagrammatic means a Bose-Fermi mixture, with boson-fermion coupling tuned by a Fano-Feshbach resonance.
For increasing coupling, the growing boson-fermion pairing correlations 
progressively reduce the boson condensation temperature and make it eventually vanish at a critical coupling. Such quantum critical point depends very weakly on the 
population imbalance and for vanishing boson
densities coincides with that found for the polaron-molecule transition in a strongly imbalanced Fermi gas, thus bridging two quite distinct physical systems.         
\end{abstract}

\pacs{03.75.Ss,03.75.Hh,32.30.Bv,74.20.-z}
\maketitle

One of the main reasons behind the recent great interest in ultracold gases, is the possibility to reproduce physical systems relevant to other areas 
in physics, with a flexibility and a degree of tunability of physical parameters which is unimaginable in the original system of interest.
%Even more exciting is, however, the possibility to use 
The very same flexibility can also be used, however, to construct novel physical systems.
%with no corresponding counterparts in other domains 
%of physics.
A noticeable example of such novel systems, namely, resonant Bose-Fermi mixtures, will be here at issue.
% we consider an ultracold gas of single-component bosons and fermions in the presence of a {\em broad} 
%Fano-Feshbach resonance, and focus on the evolution of the system .
We will be interested in particular in the competition between boson-fermion pairing correlations and bosonic condensation, occuring in a mixture of single-component bosons and fermions when the boson-fermion pairing is made
progressively stronger by means of a Fano-Feshbach resonance.  

Initial theoretical studies of ultracold Bose-Fermi mixtures  
considered mainly non-resonant systems~\cite{Viv00,Yi01,Alb02,Rot02,Lew04}, where boson-fermion 
pairing is irrelevant, and studied, within mean-field like treatments, the tendency towards collapse 
or phase separation (as motivated by the first experimental results on non-resonant Bose-Fermi mixtures~\cite{Mod02}).
Bose-Fermi mixtures in the presence of a 
Fano-Feshbach resonance have been subsequently considered in~\cite{Rad04,Kag04,Sto05,Dir05,Pow05,Avd06,Pol08,Riz08,Wat08,Mar08,Tit09}, 
mostly for {\em narrow} resonances~\cite{Rad04,Dir05,Pow05,Avd06,Mar08,Pol08} and/or in specific contexts such as 
optical lattices, reduced dimensionality, zero temperature or vanishing densitiy of one component~\cite{Kag04,Sto05,Riz08,Wat08,Tit09}. 

%For a narrow resonance one has to take into account explicitly the 
%(closed-channel) molecular state, leading to a Hamiltonian which includes 
%three different species (bosons, fermions, and molecules) from the outset. 
%Calculations with this Hamiltonian~\cite{Rad04,Dir05,Pow05,Pol08,Bor08,Tit09,Mar08,Mar09} were done by using mean-field like approximations, 
%which are justified however only in the limit of a very narrow resonance.

Most of current experiments on Bose-Fermi mixtures~\cite{Gun06,Osp06,Osp06b,Zir08,Ni08} appear however to be closer to 
the case of a {\em broad} resonance, which is characterized by the smallness of the effective range parameter $r_0$ of the 
interaction potential  
 with respect to both the average interparticle distance and the scattering length~\cite{Sim05}.
%, which is 
%anyway more appealing both for experiments, because of the practical difficulties in working with narrow resonances, 
%and in the theoretical perspective, due to the ``universal'' character of a broad resonance.
%, such that the molecular bound state results from 
%pairing of its constituent particles in the open channel.
Under these conditions, the resonant Bose-Fermi mixture is accurately described by a {\em minimal} Hamiltonian, 
made just by bosons and fermions mutually interacting via 
an attractive point-contact potential. 
This Hamiltonian is simpler and more ``fundamental'' in character than its counterpart for a narrow resonance because 
of the absence of any other parameter 
besides the strength of the boson-fermion interaction. 
In addition, its simplicity allows the implementation of more sophisticated many-body calculations (beyond mean-field).
 
We thus consider the following (grand-canonical) Hamiltonian:
\begin{eqnarray}
H&=&\sum_{s}\int\! d {\bf r} \psi^{\dagger}_s({\bf r})(-\frac{\nabla^2}{2 m_s}-\mu_s)
\psi_s({\bf r}) \nonumber\\
&+& v_0 \int\! d{\bf r} \psi^{\dagger}_B({\bf r})\psi^{\dagger}_F({\bf r})
\psi_F({\bf r})\psi_B({\bf r})
\label{hamiltonian}
\end{eqnarray}
Here $\psi^{\dagger}_s({\bf r})$, creates a particle of mass $m_s$ and chemical potential $\mu_s$ at 
spatial position ${\bf r}$, where $s=B,F$ indicates the boson and 
fermion atomic species, respectively, while $v_0$ is the bare strength
of the contact interaction (we set $\hbar=k_B=1$ throughout this paper). 
Ultraviolet divergences associated with the use of a contact 
interaction in (\ref{hamiltonian}) are eliminated, as for two-component Fermi gases~\cite{Pie00},
by expressing the bare interaction $v_0$ in terms of the boson-fermion scattering length 
$a$ via the (divergent) expression
%\begin{equation}
%\frac{1}{v_0}=\frac{m_r}{2\pi a}-\int \! \frac{d{\bf k}}{(2 \pi)^3} 
%\frac{2 m_r}{{\bf k}^2}\;,
%\end{equation}
$1/v_0= m_r/(2\pi a)-\int \! 2 m_r /{\bf k}^2 d{\bf k}/(2 \pi)^3$,
where $m_r=m_B m_F/(m_B+m_F)$ is the reduced mass. 

The Hamiltonian (\ref{hamiltonian}) does not contain explicitly the 
boson-boson interaction. Provided it is repulsive and non-resonant,
this interaction yields in fact a mean-field shift, which consists in 
a simple redefinition of the boson chemical potential 
(attractive boson-boson interactions would lead to mechanical instability and are excluded from our consideration).
S-wave interaction between fermions is finally excluded by Pauli 
principle.

The effective coupling strength in the many-body system is determined by 
comparing the boson-fermion scattering length $a$ with the average 
interparticle distance $n^{-1/3}$ (with $n=n_B+n_F$, where $n_B$ and
$n_F$ are the boson and fermion particle number density, respectively).
In particular, we will use the same dimensionless coupling strength $(k_F a)^{-1}$ normally used for two-component Fermi gases, where 
the wave-vector $k_F\equiv (3 \pi^2 n)^{1/3}$ (note that $k_F$ coincides with the 
noninteracting Fermi wave-vector $k_F^0=(6 \pi^2 n_F)^{1/3}$ only for $n_B=n_F$).
 
% Figure 1
\begin{figure}[t]
\begin{center}
\epsfxsize=7cm
\epsfbox{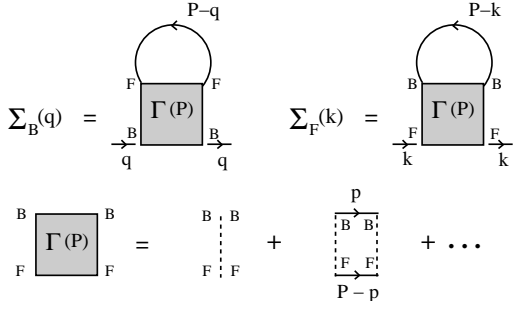}
\caption{T-matrix diagrams for the fermionic and bosonic self-energies in the normal phase. Full lines represent bare bosonic (BB) and fermionic (FF) 
Green's functions. Broken lines represent bare boson-fermion interactions 
$v_0$.} 
\end{center}
\end{figure}
The expected behavior of the system is clear in the two opposite limits 
of the boson-fermion coupling. 
In the weak-coupling limit, where the scattering length $a$ is small and negative
(such that $ (k_F a)^{-1} \ll -1$), the two components will behave essentially 
as ideal Bose and Fermi gases. Bosons will condense at 
$T_c= 3.31 n_B^{2/3}/m_B $, while fermions  will fill out (at sufficienty low temperature)
a Fermi sphere with radius $k_F^0$, the chemical potentials $\mu_B$ and $\mu_F$ being modified
with respect to their noninteracting values by the mean-field shifts $2 \pi  n_F a/m_r$, and $2 \pi n_B a/m_r$, respectively. 
In the opposite strong-coupling limit where $a$ is small and positive (such that $(k_F a)^{-1} \gg 1$ ), bosons will pair with fermions to form  
tightly bound fermionic molecules, with binding energy $\epsilon_0=1/(2 m_r a^2)$. In particular, in systems where $n_B \le n_F$, on which we focus in the present paper, all bosons will eventually pair with fermions into molecules, 
suppressing condensation completely.

In a complete analogy with the BCS-BEC crossover problem in a two-component Fermi gas~\cite{Pie00}, the choice of the self-energy diagrams 
will be guided by the criterion that a single set of diagrams should recover the correct physical description 
of the two above opposite limits. 
In the {\em normal} phase above the condensation critical temperature $T_c$, to which we restrict in the present paper,
this condition is met (as we shall see below) by the T-matrix choice 
of diagrams represented in Fig.~1, yielding the expressions:
\begin{eqnarray}\label{selfb}
\Sigma_{B}(q)&=&-\frac{1}{\beta}\int\!\!\frac{d {\bf P}}{(2\pi)^{3}}\sum_{m}G_{F}^{0}(P-q)\Gamma(P)\\
\label{selff}
\Sigma_{F}(k)&=&\frac{1}{\beta}\int\!\!\frac{d {\bf P}}{(2\pi)^{3}}\sum_{m}G_{B}^{0}(P-k)\Gamma(P)
\end{eqnarray}
for the bosonic and fermionic self-energies $\Sigma_{B}$ and 
$\Sigma_{F}$, where $G_B^0$ and $G_F^0$ are bare bosonic and fermionic Green's functions  and $\Gamma(P)$ is the many-body T-matrix: 
\begin{eqnarray}
\Gamma(P)&=&- \left\{\frac{m_{r}}{2\pi a}+\int\!\!\frac{d{\bf p}}{(2\pi)^{3}}
\right.\nonumber\\
&\times& \left.\left[\frac{1-f[\xi_{F}({\bf P}-{\bf p})]+b[\xi_{B}({\bf p})]}{\xi_{F}({\bf P}-{\bf p})+\xi_{B}\left({\bf p}\right)-i\Omega_{m}}
-\frac{2m_{r}}{{\bf p}^{2}} \right]\right\}^{-1}\!\!\!\!\!\!\!.\phantom{aa} 
\label{gamma}
\end{eqnarray}
In the above expressions $q=({\bf q},\omega_{\nu})$, $k=({\bf k},\omega_{n})$,  $P=({\bf P},\Omega_{m})$, where $\omega_{\nu}=2\pi\nu/ \beta$  and $\omega_{n}=(2n+1)\pi /\beta$, $\Omega_m=(2m+1)\pi/\beta$ 
are bosonic and fermionic Matsubara frequencies, respectively ($\beta=1/T$ being the inverse temperature and $\nu,n,m$ integer), while $f(x)$ and $b(x)$ are the Fermi and Bose distribution functions and
$\xi_s({\bf p})={\bf p}^2/(2m_s)-\mu_s$. The above self-energies determine finally the
dressed Green's functions $G_s^{-1}=G_s^{0 \; -1}-\Sigma_s$ entering the particle numbers equations:
\begin{eqnarray}\label{nb}
n_{B}&=&-\int\!\!\frac{d {\bf q}}{(2\pi)^{3}}\frac{1}{\beta}\,\sum_{\nu}G_{B}({\bf q},\omega_{\nu})\,e^{i\omega_{\nu} 0^+}\\
\label{nf}
n_{F}&=&\int\!\!\frac{d {\bf k}}{(2\pi)^{3}}\frac{1}{\beta}\,\sum_{n}G_{F}({\bf k},\omega_{n})\,e^{i\omega_{n} 0^+}\,.
\end{eqnarray}
Equations~(\ref{selfb}$-$\ref{nf}) fully determine the thermodynamic properties of the Bose-Fermi mixture in the normal phase. This phase is characterized
by the condition $\mu_{B}-\Sigma_{B}(q=0)<0$; upon lowering the temperature, condensation will then occur when $\mu_{B}=\Sigma_{B}(q=0)$.
% (signalling the start 
%of a macroscopic occupation of the zero momentum mode).  

Analytic results can be obtained in the two opposite limits of the boson-fermion coupling~\cite{details}.
For weak-coupling $\Gamma(P)\simeq -\frac{2\pi a}{m_{r}}$, yielding $\Sigma_B\simeq 2 \pi  n_F a/m_r$ and $\Sigma_F\simeq2 \pi n_B a/m_r$, in accordance with the expected mean-field shifts. 
For strong-coupling $\Gamma(P)$ gets instead proportional to the molecular propagator:
\begin{equation}
\Gamma(P)\simeq -\frac{2\pi}{m_r^2 a}\frac{1}{i\Omega_m-\frac{{\bf P}^2}{2 M}+\mu_M}
\end{equation}
where $\mu_M\equiv\mu_B+\mu_F + \epsilon_0$ is the molecular chemical potential and $M=m_B+m_F$.
Insertion of this propagator in Eqs.~(\ref{selfb}$-$\ref{nf}) leads, for $n_B < n_F$, to $\mu_M\simeq(6\pi^2 n_B)^{2/3}/2 M$, $\mu_F\simeq[6\pi^2 (n_F-n_B)]^{2/3}/2 m_F$ 
and $\mu_B=\mu_M-\mu_F - \epsilon_0\simeq - \epsilon_0$, as expected 
when all bosons pair with fermions. 
%The achievement of a sensible description of both weak and strong-coupling limits justifies then 
%the selection of the T-matrix diagrams.
% Figure 2
\begin{figure}[t]
\begin{center}
\epsfxsize=7cm
\epsfbox{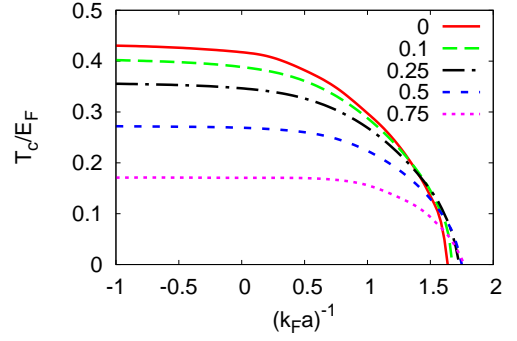}
\caption{Critical temperature (in units of $E_F=k_F^2/ 2 m_F$) for condensation of bosons vs 
the boson-fermion coupling  $(k_F a)^{-1}$ for different values of 
the density imbalance $(n_F-n_B)/n$ in a mixture with $m_B=m_F$.
% The region where the critical temperature vanishes is 
%expanded in the lower panel.
} 
\end{center}
\end{figure}

For intermediate values of the
boson-fermion coupling  
Eqs.~(\ref{selfb}$-$\ref{nf}) need to be 
solved numerically.
The results for the condensation critical temperature $T_c$ vs
the boson-fermion dimensionless coupling $(k_F a)^{-1}$  are presented in Fig.~2 for a mixture with 
$m_B=m_F$ at several  values of density imbalance.~\cite{footnotemasses}
 All curves start in weak-coupling from the corresponding 
noninteracting values, and decrease monotonically for increasing coupling due to the growing pairing correlations, which tend to deplete the 
zero momentum mode and distribute the bosons over a vast momentum region, as required to 
build the internal molecular wave-function. The critical temperature vanishes 
eventually at a critical coupling, corresponding to a quantum critical
point which separates a phase with a condensate from a phase where molecular 
correlations are so strong to deplete the condensate completely.

Remarkably, the critical coupling value depends very weakly on the degree of density 
imbalance: all curves terminate in the narrow region $1.6<(k_F a)^{-1}<1.8$. 
In this respect, it is interesting to consider the limit $n_B\to 0$, where the critical coupling can be 
calculated independently by solving the problem of a single boson immersed in a Fermi sea. This 
is actually the same as a spin-down fermion immersed in a Fermi sea of spin-up fermions, since
for a single particle the type of statistics is immaterial.  
The critical coupling reduces then to that for the polaron-to-molecule transition, 
recently studied in the context of strongly imbalanced {\em two-component Fermi gases}~\cite{sch09,vei08,pro08,mas08,mor09,pun09,com09}.
In particular, the solution of our equations in this limit yields the critical coupling $(k_F a)^{-1}=1.60$, 
in full agreement with the value $(k_F^{\uparrow} a)^{-1}=1.27$ (where $k_F^{\uparrow}=2^{1/3} k_F$) reported in~\cite{pun09,com09}  for   
the polaron-to-molecule transition, when treated at the same level of approximation~\cite{footnote-a,footnote-b}. 
The meeting of the properties of a Bose-Fermi mixture with those of a two-component Fermi mixture at this point of the phase diagram is
quite remarkable, especially because, according to our results of Fig.~1, this ``universal'' point sets the scale for the quantum phase 
transition for {\em all} boson densities $n_B\le n_F$.

% Figure 3
\begin{figure}[t]
\begin{center}
\epsfxsize=7cm
\epsfbox{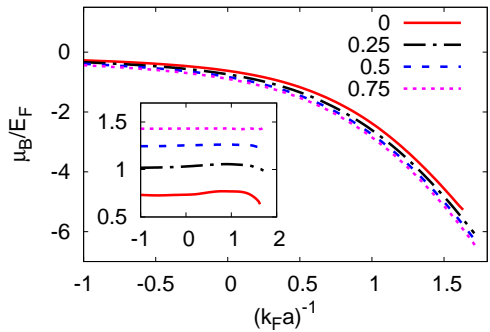}
\caption{Boson chemical potential at the critical temperature $T_c$ (in units of $E_F$) vs the boson-fermion coupling $(k_F a)^{-1}$ for different
values of the density imbalance $(n_F-n_B)/n$ in a mixture with $m_B=m_F$. The 
corresponding fermion chemical potential is reported in the inset.} 
\end{center}
\end{figure}
The chemical potentials $\mu_B$ and $\mu_F$ at $T_c$ vs.~$(k_F a)^{-1}$ are reported in Fig.~3, for different
values of the density imbalance. The two chemical potentials behave 
quite differently. The fermion chemical potential remains almost constant in the whole range of coupling considered; the boson chemical
potential, on the other hand, diminishes quite rapidly with increasing coupling while depending little on the density imbalance. This different
behavior results from the concurrence of several factors. For weak-coupling, the increasing (negative) mean-field shift of the fermion chemical potential for
increasing coupling is partially compensated by the decrease of the temperature when moving along the critical line.
On the molecular side, the fermion chemical potential is instead determined by the Fermi energy of the unpaired fermions plus a mean-field shift
caused by interaction with molecules. Pauli repulsion makes this interaction repulsive~\cite{details,Kag04} thus keeping the fermion chemical potential positive. 
%These constraints determine the small range of variations of the fermion chemical potential along the critical line and its dependence on
%the density imbalance.
The boson chemical potential, on the other hand, interpolates between the mean-field value  $2 \pi  n_F a/m_r$ in weak coupling and $\mu_B\approx -\epsilon_0$ 
in strong coupling, as required by molecule formation.
% Figure 4
\begin{figure}[t]
\begin{center}
\epsfxsize=7cm
\epsfbox{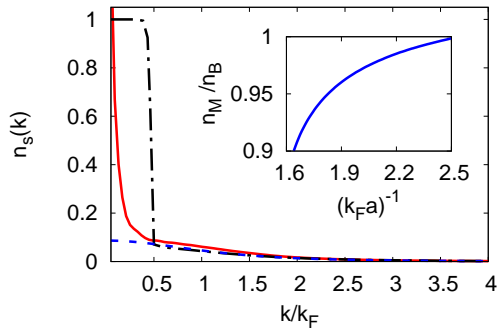}
\caption{Boson (full line) and fermion (dashed-dotted) momenta distribution 
curves for $n_B=n_F$, $m_B=m_F$, $(k_F a)^{-1}=1.63$ and 
$T=T_c\simeq 0.015 E_F$.
See text for the meaning of the dashed curve. The inset reports the fraction of molecules vs $(k_F a)^{-1}$ for the same mixture at 
$T=0.1 E_F$.} 
\end{center}
\end{figure}

Figure 4 reports the momentum distributions $n_B({\bf k})$ and $n_F({\bf k})$ (as obtained 
before momentum integration in Eqs.~(\ref{nb})-(\ref{nf})) at $T_c$, for a mixture with $n_B=n_F$, at
the coupling value $(k_F a)^{-1}=1.63$ (approximately at the quantum critical point). 
 The two distributions are markedly different at low momenta, consistently 
with their different statistics, but coalesce 
into the same behavior just after the step in the fermion momentum distribution. 
%(namely, for $k\gtrsim 0.5 k_F$ for the case considered here).
This common behavior  corresponds to the function $n_M |\phi({\bf k})|^2$ (dashed line in Fig.~4), where 
$\phi({\bf k})=(8\pi a^3)^{1/2}/({\bf k}^2 a^2+1)$ is the normalized two-body internal wave function
of the molecules, while the coefficient $n_M$ represents 
their density. In the present case $n_M=0.89 \, n_B$, showing that 
a fraction of bosons remains unpaired but still does not condense even at such a low temperature.
The extrapolation of these results at exactly zero temperature indicates the existence (in a coupling range starting right after the quantum critical 
point) of quite an unconventional Bose liquid, corresponding to the unpaired 
bosons, which do not condense even at zero temperature. 
The fraction of unpaired fermions is instead more conventional
and consists in a Fermi liquid, which is responsible for the jump in the fermion momentum 
distribution. In particular, the position of the jump in Fig.~4 at 
 $|{\bf k}|\simeq0.47 \, k_F$,  
corresponds to an ``enclosed'' density of $0.10  \, n_F$, in good numerical 
agreement with the value $0.11 \, n_F$ obtained from the Luttinger theorem~\cite{lut60} 
for the fraction ($n_F-n_M$) of unpaired fermions (using the value $n_M=0.89\, n_B$ 
extracted independently above). 
Note finally that the number of unpaired bosons progressively 
decreases by increasing the coupling, as expected on physical grounds, 
reaching eventually a 100\% conversion of bosons into molecules, 
as shown in the inset of Fig.~4, where the ratio $n_M/n_B$ is reported vs coupling at a constant temperature.

We wish to conclude finally by commenting on three-body losses and 
mechanical instabilities, that could prevent the experimental observation
of the many-body physics described above.
The choice of restricting our analysis to mixtures with a majority of fermions was precisely aimed at
reducing the importance of these nuisances.
Three-body losses are dominated by processes involving two bosons and one fermion, with a rate
proportional to $n_B^2 n_F$~\cite{Zir08}. They can thus be controlled
by keeping $n_B$ sufficiently small with respect to $n_F$.
A predominance of fermions should also reduce the tendency to collapse, because
of the stabilizing effect of Fermi pressure. As a matter of fact, we have 
verified that the compressibility matrix $\partial\mu_s/\partial n_{s'}$ 
remained positive definite for all couplings, temperatures, and densities 
$n_F\ge n_B$ considered in our calculations.
Apparently, pairing correlations act to protect the system from the mean field 
instabilities dominating the phase diagram of non-resonant 
Bose-Fermi mixtures. Resonant Bose-Fermi mixtures with a majority of 
fermions appear thus promising systems for the observation of a rich  
many-body physics. 

\acknowledgments
We thank F.~Palestini, A.~Perali, and G.C.~Strinati for a careful reading of the manuscript. Partial support by the Italian MIUR under Contract Cofin-2007 ``Ultracold atoms and novel quantum phases'' is acknowledged.


\begin{thebibliography}{99}
\bibitem{footnotemasses} Numerical calculations will be limited to the case $m_B=m_F$ to restrict the
space parameter. This case is directly applicable to isotopic mixtures of 
sufficiently heavy atoms ($^{39}$K-$^{40}$K, e.g.), but it is taken as 
representative of the more general situation.
\bibitem{footnote-a}
QMC calculations~\cite{pro08} and refined diagrammatic approximations~\cite{com09} for the
single spin-down problem yield the value $(k_F a)^{-1}=1.11$ 
for the polaron-to-molecule transition, a
30\%  off the T-matrix prediction.
This difference is due to the overestimate of the molecule-fermion repulsion by the T-matrix approximation,
which yields $8/3 a$ for the molecule-fermion scattering length in place of the exact value $1.18 a$~\cite{sko56}, 
thus making the molecule formation in a Fermi sea enviroment less convenient.
\bibitem{footnote-b}
The value 1.60 for the critical coupling in the limit $n_B\to 0$ is reached with a weak reentrant behavior of the critical coupling vs.~imbalance, 
occurring at imbalances larger than those reported in Fig.~1. For instance, at imbalance 0.95, $T_c$ vanishes at $(k_F a)^{-1}\simeq1.63$.
\end{thebibliography}
\end{document}